\documentclass[prb,aps,twocolumn,superscriptaddress,10pt,showpacs,longbibliography]{revtex4-1}
\usepackage{graphicx}% Include figure files
\usepackage{epstopdf}
\usepackage{color}
\usepackage{amsmath,amssymb}
\usepackage{fixmath}

\usepackage [english]{babel}
\usepackage [autostyle, english = american]{csquotes}
\MakeOuterQuote{"}

\begin{document}

\title{Improper Ferroelectric Polarisation in a Perovskite driven by Inter-site Charge Transfer and Ordering}

\author{Wei-Tin Chen} 
\affiliation{Center for Condensed Matter Sciences, National Taiwan University, Taipei 10617, Taiwan}
\affiliation{Taiwan Consortium of Emergent Crystalline Materials, Ministry of Science and Technology, Taipei 10622, Taiwan}

\author{Chin-Wei Wang}
\affiliation{National Synchrotron Radiation Research Center, Hsinchu 30076, Taiwan}

\author{Hung-Cheng Wu}
\affiliation{Department of Physics, National Sun Yat-Sen University, Kaohsiung 80424, Taiwan}

\author{Fang-Cheng Chou}
\affiliation{Center for Condensed Matter Sciences, National Taiwan University, Taipei 10617, Taiwan}
\affiliation{Taiwan Consortium of Emergent Crystalline Materials, Ministry of Science and Technology, Taipei 10622, Taiwan}
\affiliation{National Synchrotron Radiation Research Center, Hsinchu 30076, Taiwan}

\author{Hung-Duen Yang}
\affiliation{Department of Physics, National Sun Yat-Sen University, Kaohsiung 80424, Taiwan}

\author{Arkadiy Simonov}
\affiliation{Department of Chemistry, University of Oxford, South Parks Road, Oxford OX1 3QR, United Kingdom}

\author{M. S. Senn}
\email{m.senn@warwick.ac.uk}
\affiliation{Department of Chemistry, University of Warwick, Gibbet Hill, Coventry, CV4 7AL,United Kingdom}

\date{\today}

\begin{abstract}

It is of great interest to design and make materials in which ferroelectric polarisation is coupled to other order parameters such as lattice, magnetic and electronic instabilities.  Such materials will be invaluable in next-generation data storage devices.   Recently, remarkable progress has been made in understanding improper ferroelectric coupling mechanisms that arise from lattice and magnetic instabilities.  However, although theoretically predicted, a compact lattice coupling between electronic and ferroelectric (polar) instabilities has yet to be realised.  Here we report detailed crystallographic studies of a novel perovskite Hg$^{\textbf{A}}$Mn$^{\textbf{A'}}_{3}$Mn$^{\textbf{B}}_{4}$O$_{12}$ that is found to exhibit a polar ground state on account of such couplings that arise from charge and orbital ordering on both the A' and B-sites, which are themselves driven by a highly unusual Mn$^{A'}$-Mn$^B$ inter-site charge transfer.  The inherent coupling of polar, charge, orbital and hence magnetic degrees of freedom, make this a system of great fundamental interest, and demonstrating ferroelectric switching in this and a host of recently reported hybrid improper ferroelectrics remains a substantial challenge.

\end{abstract}


\maketitle

\section{Introduction}
\begin{figure*}[t]
\includegraphics[width=18.3cm] {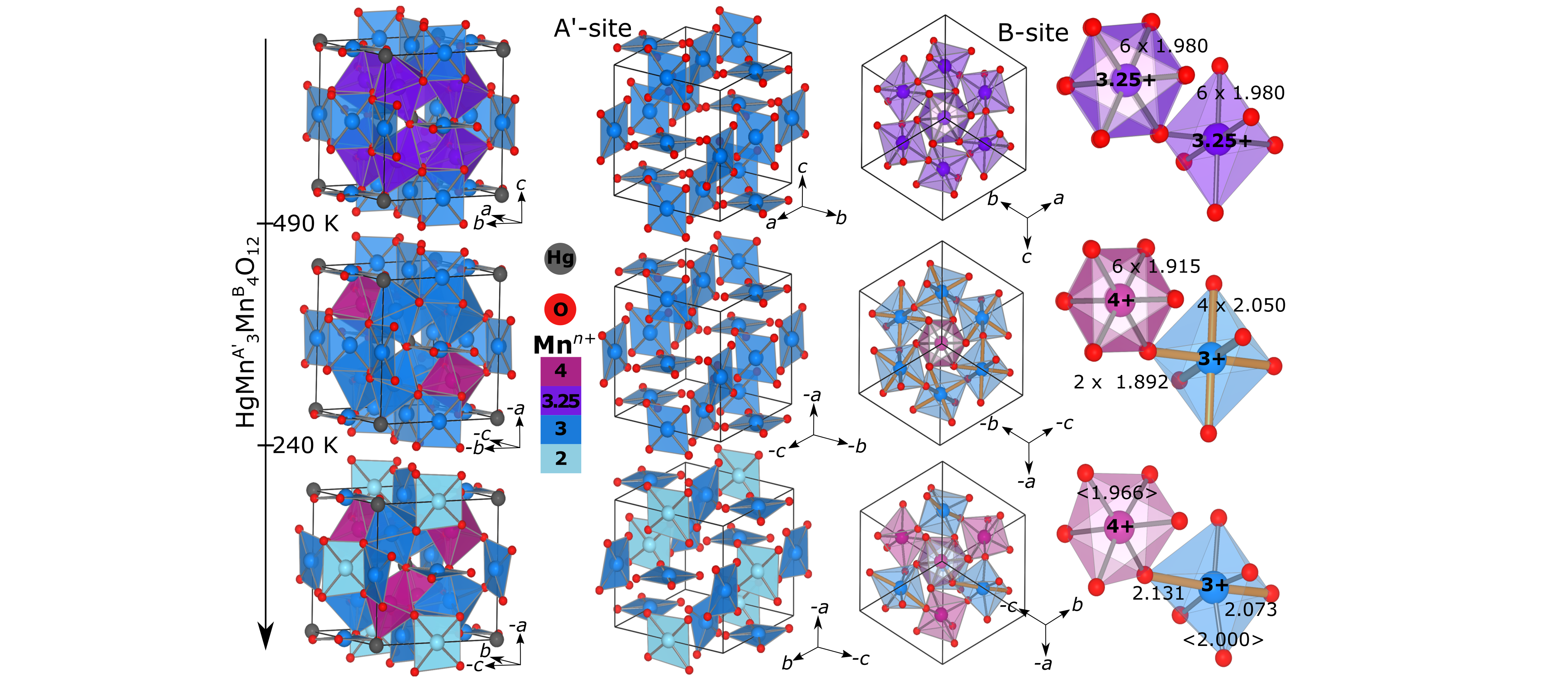}
\caption{\label{structure} The structure of the phases of HgMn$_{7}$O$_{12}$.  Top panel, high temperature cubic aristotype $Im\bar{3}$; middle, $R\bar{3}$ orbital disorder phase; bottom, $Pnn2$, double charge ordered, polar phase.  The colour key indicates the formal valence states of Mn sites assigned based on analysis of the crystal structure. The orange bonds represent the Jahn-Teller long bonds (d$_z$ orbital order) of the Mn$^{3+}$. The $R\bar{3}$ phase has 4 apparent Jahn-Teller long bonds in the unit cell which are actually an average of 2-short and 2-long bonds, observed on account of the orbital disorder.  There is a charge transfer of one electron per unit cell from B to A' on going from $R\bar{3}$ to $Pnn2$ which acts to remove the orbital disordered state and facilitates the double charge order on the A' and B-sites.}
\end{figure*}

Considerable efforts have been made in the search for novel multiferroics, as such materials will undoubtedly find utility in future solid state storage devices, driving both capacity and data access speeds ever higher. In these materials, it is essential to investigate the coupling between ferroelectric polarisation and other order parameters.\cite{Bousquet2008, Benedek2011,Eerenstein2006,Efremov2004a, VandenBrink2008a} Perovskites are a family of materials exhibiting a large number of technologically desirable physical properties.  However, despite their very rich chemistry, ferroelectricity has been restricted to a rather narrow range of compositions, and there is good theoretical evidence to suggest that anti-polar displacive instabilities generally act to suppress polar ones \cite{Benedek2013,Aschauer2014}. For the most part, experimental studies have hence been confined to Ti, Pb of Bi containing perovskites, where a strong desire for these cations to undergo a second-order Jahn-Teller distortion provides a local driving force for the resulting proper ferroelectricity.  On the other hand, in the field of improper ferroelectrics, remarkable progress has been made in understanding the coupling mechanisms that arise from lattice and magnetic instabilities\cite{Harris2007,Barone2015,Benedek2015}. For instance, recent work in the closely related layered Ruddlesden-Popper perovskites have led to theoretical predictions of improper ferroelectricity arising due to the coupling of two tilt modes of BO$_{6}$ octohedra \cite{Benedek2011} to a third polar instability.  However, for un-layered perovskites, it would appear that structural symmetry conspires to make such a coupling scheme utilising tilt modes alone, impossible. Furthermore, despite theoretical predictions,~\cite{Bristowe2015, Varignon2015,Stroppa2013,Yamauchi2013} a compact lattice coupling between electronic (orbital and charge degrees of freedom) and ferroelectric (polar) instabilities has yet to be experimentally realised. On the other hand, some perovskite materials are known to exhibit improper-ferroelectricity by a so called type-II multiferroic mechanism, in which the polarisation is a secondary order parameter of the primary magnetic ordering~\cite{Khomskii2009a}. One of the most notable examples is CaMn$_7$O$_{12}$ which has a so-called 134-Perovskite AA'$_3$B$_4$O$_{12}$ structure (top panel Fig.~\ref{structure} for aristotype). This example was believed to represent the largest magnetically induced polarisation reported to date ~\cite{Zhang2011,Johnson2012}, although more recent findings dispute this ~\cite{Terada2016}.  This magnetoelectric has been found to be associated with a magnetic ordering arising due to an unusual helical orbital order~\cite{Perks2012}.  Despite the significant magnetoelectric response in this material, the ordering temperature remains low, as the mechanism still explicitly relies on the magnetic ordering, rather than just the orbital ordering which occurs at much higher temperatures.  It is hence desirable to explore chemistries around CaMn$_{7}$O$_{12}$ and associated local degrees of freedom with the hope of coupling orbitally ordered states directly to ferroelectric properties.  With the above in mind, we have synthesised the novel perovskite HgMn$_7$O$_{12}$ using high pressure synthesis techniques. 

\section{Experimental Details and Data Analysis}

\subsection{Sample Preparation}
Polycrystalline HgMn$_{7}$O$_{12}$ was prepared by solid-state reaction under high pressure and high temperature conditions. Stoichiometric amounts of HgO (Sigma-Aldrich $\geq$ 99.0\%), MnO$_2$ (Alfa Aesar 99.997\%) and Mn$_2$O$_3$ (Aldrich 99.99\%) were well mixed and sealed in a platinum capsule. The capsule, boron nitrite insulating layer and graphite heater were assembled in a pyrophyllite cell and placed in a DIA-type cubic anvil high-pressure apparatus. The sample was treated at 8 GPa and 1600 K for 30 min then released to ambient condition. Each synthesis produced $\approx$ 0.1 g sample. Judging by X-ray powder diffraction and magnetic susceptibility (See S.I., Fig. S1), products from 14 out of the 24 runs were selected and combined for further analysis in the neutron powder diffraction studies.

\subsection{Crystallography and Crystal Structure Analysis}

Full details of the refined average structures are given in the the S.I. (Table S1). Synchrotron X-ray powder diffraction (SXRD) experiments were carried out for phase identification and crystal structure analysis. The powder sample was packed in a 0.1 mm borosilicate capillary tube to minimize the absorption effect and measured with a 15 keV beam and the MYTHEN detector at beamline 09A, Taiwan Photon Source and I11, Diamond Light Source, UK. Neutron powder diffraction data on the combined 1.5 g polycrystalline sample was carried out at beamline WOMBAT~\cite{Studer2006} and ECHIDNA~\cite{Liss2006}, ANSTO, Australia. Temperature dependent diffraction patterns were collected in a 6 mm diameter vanadium can and a neutron wavelengths of  1.63 \AA ~and 2.42 \AA. The obtained data were analysed within the Rietveld method using the program \textit{TOPAS}, and refinements were done using the symmetry adapted displacements formalism as parametrised by $ISODISTORT$\cite{Campbell2006} and implemented through the $Jedit$ interface with \textit{TOPAS}~\cite{Campbell2007}. The parametrisations were all performed with respect to our high temperature cubic structure ($Im\bar{3}$, (550 K)), with a setting where Hg sits on (0,0,0); A' Mn1 (0,0.5,0.5); B Mn2 (0.25, 0.25, 0.25) and O ($x$, $y$, 0).  For the rhombohedral phase (room temperature), in order to facilitate an easy comparison between this, the parent structure and the orthorhombic phase, we adopted here a $R\bar{3}$ rhombohedral setting which retains the parent unit cell axes ($a$ = 7.37987 \AA, $\alpha$ = 90.37922, [(0,0,-1),(0,-1,0),(-1,0,0)], origin=(0,0,0)) applying the appropriate symmetry constraints. 

Careful examination of the neutron diffraction data at 100 K revealed additional peaks which violate the body-centring of the parent $Im\bar{3}$ cell (Fig. \ref{data} (d)). Only very weak signatures of these additional reflections could be found in the high resolution synchrotron X-ray diffraction data (Fig. \ref{data} (e)), indicating that they most likely arise due to subtle distortions of the oxygen atoms. In the propagation vector formalism these additional weak reflections can be indexed with k = [1,1,1] (H-point).  In a similar spirit to the exhaustive approach for structural determination employed recently in the case of the phase transitions in Bi$_{2}$Sn$_{2}$O$_{7}$~\cite{Lewis2016}, we generated a complete list of possible structural models which are necessarily subgroups of the parent symmetry $Im\bar{3}$ and are compatible with both the observations of metric orthorhombic symmetry, and the occurrence of reflections that can be indexed on k = [1,1,1] .  In total we tested 10 models and, while many gave near equivalent fitting statistics to the combined neutron and synchrotron X-ray diffraction Rietveld refinements that we performed on data collected at 100 K, only the $Pnnm$ and $Pnn2$ models provided a satisfactory fit to the weak superstructure reflections (See Fig. \ref{data} (d) and (e)). To facilitate a discrimination between these models, several grains of $\approx$ 10 micron diameter were selected from the powder for single crystal diffraction data collections at 100 - 120 K.  The single crystal data was measured on the I19 beamline, Diamond Light Source equipped with a Pilatus 2M detector. The measurements were performed with a wavelength of 0.6889 \AA. Single crystal Bragg intensities were integrated using the program $XDS$~\cite{Kabsch2010}. Since $XDS$ does not allow multirun experiments to be processed, all runs were integrated separately by manually transforming the crystal orientation matrix and then manually scaling them together. A combined refinement in $TOPAS$ against this single crystal data and the neutron and X-ray powder diffraction data, clearly show that the polar space group $Pnn2$ ($a$ = 7.22271(3) \AA, $b$ = 7.32475(3) \AA,$c$ = 7.58504(3) \AA,[(0,1,0),(0,0,-1),(-1,0,0)], origin=(0,0,0)) provides a significantly better fit to the data (see S.I., Table S2, S3 and Fig. S2). The reciprocal space reconstructions presented here (Fig. \ref{S4} and \ref{S5}) are from two different measurements taken at 120 K.

%The atomic displacements arising due to the charge and orbital order on the B-sites in the $Pnn2$ phase were decomposed in the web-based program $ISODISTORT$ ~\cite{Campbell2006} as transforming as the irreducible representations H$_{1}^{-}$ and H$_{2}^{-}$H$_{3}^{-}$ of the parent $Im\bar{3}$ phase.  A single order parameter transforming as H$_{2}^{-}$H$_{3}^{-}$ is sufficient to induce the observed orthorhombic distortion and drive the aforementioned charge and orbital order on the B-site.  However, the action of this order parameter lowers the space group symmetry only as far as $Pnnn$ and an additional primary order parameter transforming as H$_{4}^{+}$ is required to lower the space group symmetry to the observed $Pnn2$.  The role of the additional degrees of freedom in the $Pnn2$ model (that transform as H$_{4}^{+}$) is to help restore the orbital degeneracy on the A'-site Mn$^{2+}$ by making the coordination environment more regular and shortening the second nearest neighbor distance (see Fig. \ref{polarisation}).  The tilinear term responsible for the improper ferroelectric polarisation discussed in this letter is H$_{2}^{-}$H$_{3}^{-}$  H$_{4}^{+}$ P.

\subsection{Magnetic and Transport Property Measurements}

Magnetic property measurements were carried out with a commercial magnetometer Quantum Design Magnetic Properties Measurement System (VSM-MPMS).  Zero field cooled and field cooled magnetic susceptibility were measured at 5-300 K in an external magnetic field of 10 kOe. The dielectric constant measurements were made using the Agilent 4294A precision impedance analyser by applying an AC voltage of 1 V. For pyroelectric current measurements, an electrical poling field was applied during the cooling down to base-temperature and pyroelectric current was obtained upon heating using a Keithley 6517B electrometer.

Temperature dependent dielectric constant measurements of HgMn$_{7}$O$_{12}$ with a range of frequencies from 10 kHz to 1 MHz are presented in Fig.~\ref{data} (b), exhibiting a frequency-independent step-like anomaly near the structural transition. The long-range transition ordering represents a thermal hysteresis behaviour, which could be associated with magnetization curves. Based on the Rietveld analysis of high-resolution synchrotron X-ray patterns, HgMn$_{7}$O$_{12}$ undergoes a structural transition from centrosymmetric $R\bar{3}$ to non-centrosymmetric $Pnn2$ with a polar structure, which invokes charge ordering between the Mn A-site and the Mn B-site. The results suggest that the frequency-independent step-like anomaly could be an improper ferroelectric-like ordering. In addition, there is also a frequency-independent peak close to T$_N$. For investigating the possibility of an additional ferroelectric component in HgMn$_{7}$O$_{12}$ at lower temperatures, temperature dependent pyrocurrent measurements with different applied electric fields and heating rates were performed (see Fig. \ref{S6}). The reliable pyrocurrent peaks with different applied electric fields 0, 1, and 2 kV/cm and different heating rates 5, 7.5, and 10 per min are displayed in Fig. \ref{S6} (b). A switchable pyrocurrent peak near T$_N$ through a poling electric field of ±  1 kV/cm is represented in inset of Fig. \ref{S6}, signifying the occurrence of an additional ferroelectric component below T$_N$. 

\section{Results and Discussion}

\subsection{Structural Phase Transitions}

\begin{figure}
\includegraphics[width=8.9cm] {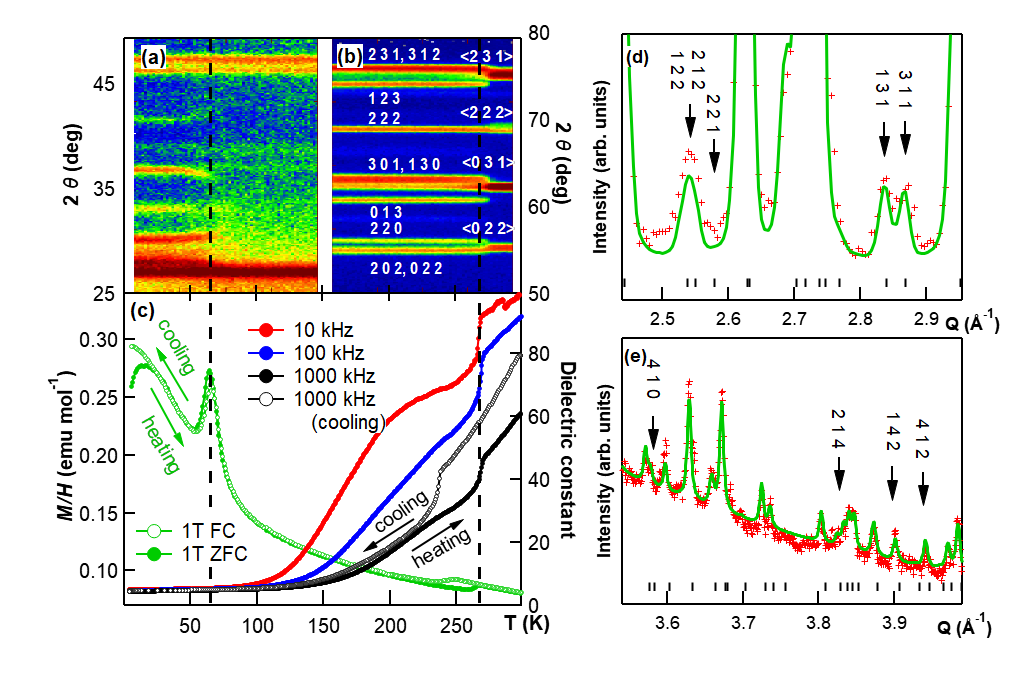}
\caption{\label{data} Structural and physical property characterisation of HgMn$_{7}$O$_{12}$. (a) and (b) show selected peaks from the neutron diffraction as a heat map, showing both the pronounced lattice distortion which occurs at 260 K on warming (240 K on cooling), and the low temperature magnetic phase transition at 60 K; (c), field and zero field cooled magnetic susceptibility taken with a measuring field of 1 T and a dielectric susceptibility measured at three different frequencies as indicated; (d) and (e) weak superstructure reflections fitted with the $Pnn2$ model.}
\end{figure}

\begin{figure}[t]
\includegraphics[width=8.9cm] {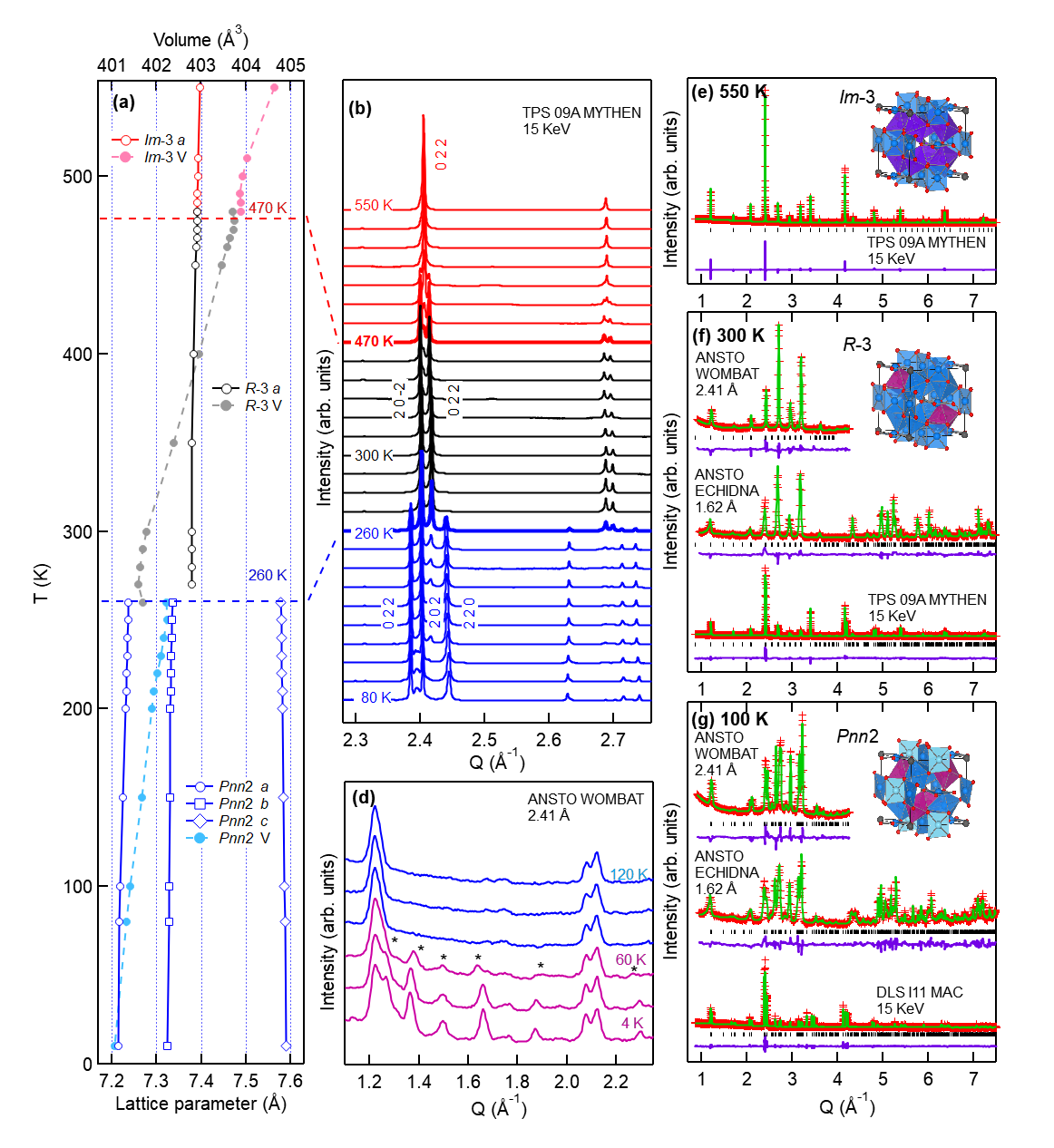}
\caption{\label{S2} (a) Temperature evolution of lattice parameters of HgMn7O12 phases. (b) and (c) Selected Q range of SXRD data showing the lattice splitting and phase evolution of HgMn7O12. (d) Selected Q range of NPD data showing the development of long range magnetic ordering. (e)-(g) Joint Rietveld refinement results of SXRD and NPD data. The insets show the crystal structure of HgMn7O12 phases.}
\end{figure}

Our variable temperature neutron and synchrotron X-ray, powder diffraction and physical property measurements (Fig. ~\ref{data}) reveal three distinct structural phase transitions, that we will focus on in this paper.  On cooling an additional low temperature (63 K) magnetic transition was observed, showing an incommensurate magnetic ordering nature. The magnetic susceptibility and isothermal magnetization results indicate a soft ferrimagnetic behavior of the compound. The details of the modulated magnetic structure will be discussed further elsewhere. Above 490 K the $Im\bar{3}$ aristotype is observed (Fig. \ref{structure} (a)). On cooling, a first order phase transition is evident via a phase coexistence in the high temperature SXRD data  between 470 - 490 K (Fig. \ref{S2}), and the structure can be indexed on the same rhombohedral space group reported for the charge ordered room temperature structure of CaMn$_{7}$O$_{12}$ ~\cite{Przenioslo2002}. 

At room temperature, we find a 1:3 charge ordered state of the Mn B-sites, with bond valence sums (BVS) extracted from our Rietveld refinements of 3.63+ : 3.24+, consistent with  previous reports for CaMn$_{7}$O$_{12}$~\cite{Przenioslo2002}.  We consider the formal valence states to be 4+~and 3+, and the magnitude of the charge segregation as observed by BVS calculations are typical of those commonly found in other metal oxides~\cite{Attfield2006}.  Further analysis of the crystal structure reveals that the Mn$^{3+}$ sites exhibit the unusual Jahn-Teller compression (2-short, 4-long Mn-O bonds) rather than elongation (2-long, 4-short) expected for Mn$^{3+}$ (t$_{2g}^3$ e$_g^1$) in an octahedral crystal field.  Invariably, such observations can be attributed to orbital disorder ~\cite{Streltsov2014}.  Here, we attribute the two short bonds being ordered along the unique rhombohedral axis, with a disordering between 2-short and 2-long bonds in the perpendicular plane (Fig. ~\ref{structure}) leading to a scenario where in the average crystallographic structure 2-short and 4-medium bond lengths are observed.

On cooling below 240 K, another crystallographic phase transition is evident by a dramatic orthorhombic lattice distortion (Fig.~\ref{data} (b)). Hysteresis in both the magnetic and dielectric susceptibilities (Fig.~\ref{data} (c)) indicate that the phase transition is strongly first order.  The large change in the dielectric constant suggests a substantial structural change has occurred at the phase transition.   Of the 10 possible orthorhombic subgroups of $Im\bar{3}$ consistent with the observed propagation vector k = [1,1,1] (H-point), only a model in $Pnn2$ provided a satisfactory fit simultaneously to the weak superstructure reflections in the powder diffraction data (Fig. \ref{data} (d) and (e))) and the single crystal data collected on grains taken from the polycrystalline samples (Fig. \ref{S4}). 

%Reciprocal space reconstruction from data collected also reveal some very weak violations of the n-glides planes suggesting that the true average symmetry can be no higher than $P112$, but any further attempt to lower the symmetry of our model did not significantly improve the fitting statistic (S.I. Table S4), and returns a model which would leave the physical interpretation presented below unchanged.  

\subsection{Diffuse Scattering and Structural Modulations}

Additional features in the single crystal diffraction data, some that evolve over time below 240 K, reveal that further very small modulation to the average crystallographic structure presented above, may occur as a competing kinetic phase.  Although our discussion in the rest of this paper will focus on the average crystallographic results, we present here for completeness all additional observed features, no matter however subtle they may be.

\begin{figure}[t]
\includegraphics[width=8.9cm] {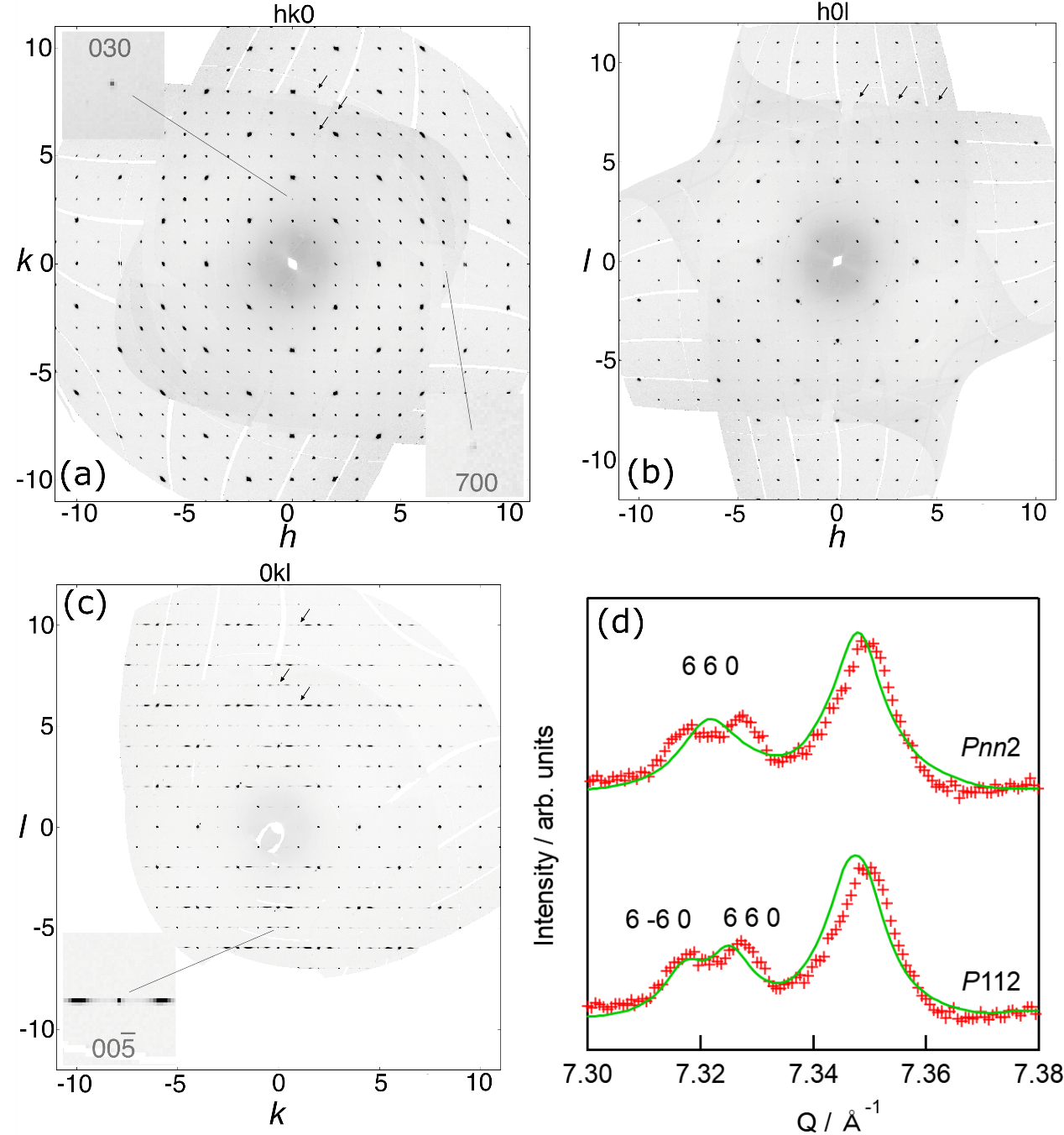}
\caption{\label{S4}  Planes of reciprocal space reconstructed from small (radius ~ 5 microns) crystal of HgMn$_7$O$_{12}$ at 100 K from data taken at I19, Diamond Light Source, showing violations of systematic absence conditions, (a) $hk0$, (b) $h0l$ and (c) $0kl$, and I11 synchrotron powder diffraction data (d) evidencing a slight monoclinic splitting.}
\end{figure}

Fig. \ref{S4} shows the hk0, h0l and 0kl planes of reciprocal space reconstructed from small (radius ~ 5 microns) crystal of HgMn$_7$O$_12$ at 100 K from data taken at I19, Diamond Light Source, showing violations of systematic absence conditions imposed by the n-glide (arrows) and 2$_1$ screw axes (insets). The violations are most evident for $P11n$ (indicated by arrows in Fig. \ref{S4} (a)) in the $hk0$ plane, and hence our initial assignment of space group $Pnnm$ or $Pnn2$ symmetry based on the powder diffraction data alone can be explained. Refinements against the integrated structure factors extracted against this data and that collected on several other crystals, reveals that the $Pnn2$ space group provides a significantly better fit to the data than centrosymmetric $Pnnm$ model and this is the bases of our final assignment of the average structure discussed above (See S.I. Fig. S2, and Tables S2 and S3).  Additional further violations of $P1n1$ and $Pn11$ glides are indicated by arrows in Fig. \ref{S4} (b) and (c), but these are at least an order of magnitude weaker than the $P11n$ violations (which are themselves already very weak).  The insets also show violation for the $P12_11$ axes and exceptionally weak violations $P2_111$, which taken together with the much stronger $P112_1$ violations (inset of Fig. \ref{S4} (c)) would suggest a space group of $P222$ or $Pmmm$.  However, neither of these are subgroup of $Pnn2$ or $Pnnm$ which were the only orthorhombic models providing a satisfactory fit to the powder diffraction data, and we must hence conclude that the true average crystallographic symmetry is no higher than $P112$ in the present setting.  There is some evidence for splitting of the high angle peaks ($h$ + $k$ $\geq$ 10) of the I11 synchrotron powder diffraction data Fig. \ref{S4}(d) supporting a $P112$ model, but we estimate that this splitting is less than $\gamma$ $<$ 90.06 $^{\circ}$. Refinements in this space group neither significantly improve the fitting statistic (see S.I., Table S4), nor change the physical interpretation of the model, and so our analysis in this paper focuses on the $Pnn2$ model.   

\begin{figure}[t]
\includegraphics[width=8.9cm] {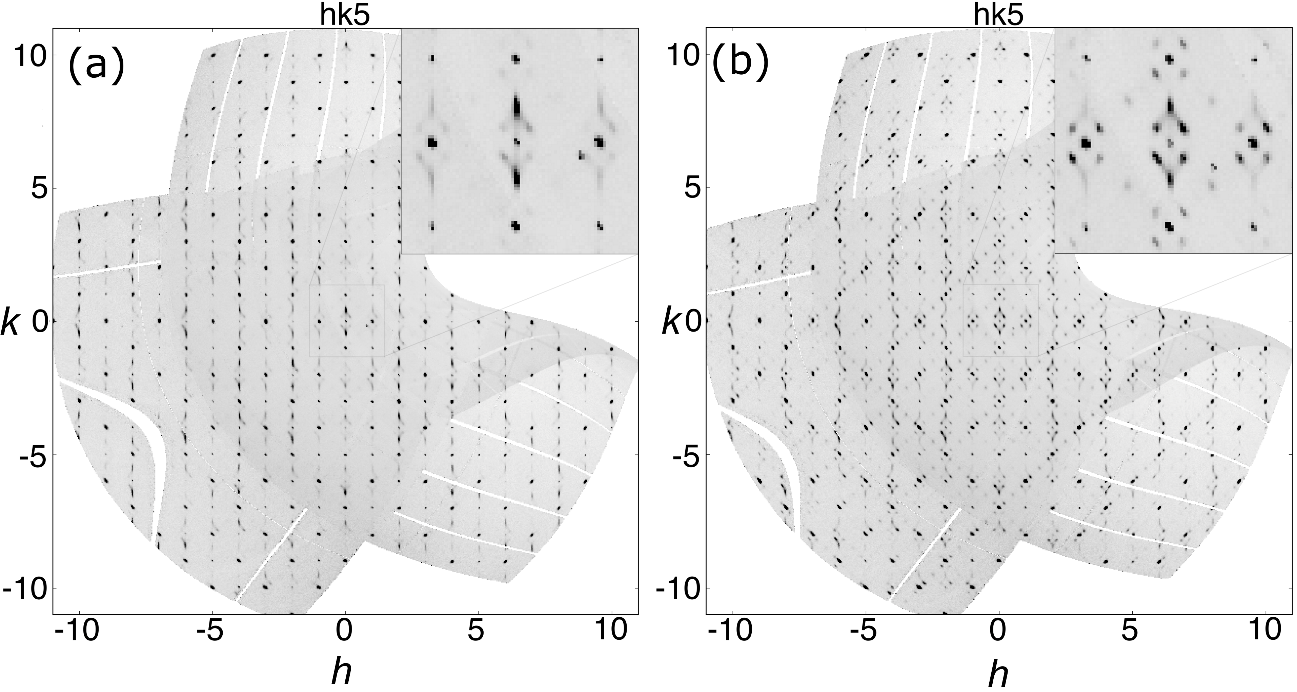}
\caption{\label{S5} Evolution of diffuse scattering below the phase transition at 260K, here shown at 120K from overexposed diffraction data, (a) reconstruction on a quickly cooled crystal, and (b), after 80 minutes at 120 K.}
\end{figure}

Evolution of diffuse scattering below the phase transition at 260 K is shown at 120 K from overexposed diffraction data in Fig. \ref{S5} (a) reconstruction on a quickly cooled crystal, and after 80 minutes at 120 K in Fig. \ref{S5}(b).  In addition to the Bragg peaks violating the I-centering condition $h+k+l$ = even, discussed above, these patterns show some other features. Right after the phase transition, the crystals develops additional streaks of diffuse scattering along the $b$* axis which can also be seen in the inset of Fig. \ref{S4} (c).  The intensity of the streaks occurs with maxima at approx. $h$, $k$ ± 0.53, $l$ (see insets in Fig. \ref{S5}).  The intensity of this diffuse scattering gradually decreases as $l$ tends to zero, and is absent from the $hk0$ plane. This is very characteristic for crystals with displacive disorder. In our case it means that part of the motion which is captured by the atomic displacement parameters has static rather than dynamic origin. In real structures, these displacements forms a highly correlated pattern in ac plane, while along the c axis the correlations decay away very quickly. Additionally, the fact that the maxima of diffuse scattering lies at a low symmetry position in reciprocal space, means that the local order along the b-axis has an incommensurate nature.  In addition to the diffuse streaks, at low temperature after some time, further features in the diffraction pattern develop. These can be characterised as satellite reflections with the modulation vectors k = [$\frac{1}{6}$, $\frac{1}{6}$, 0] and [$\frac{1}{6}$, -$\frac{1}{6}$, 0] and another set of streaks which form a Y-shaped pattern about these with the streaks along the b* axis (see insets). A possible interpretation of these features would be that the crystal is slowly turning to a commensurately modulated lower temperature phase and that the diffuse scattering streaks are formed by the domain walls between the disordered phase and low temperature modulated phase. The significance of diffuse scattering and disorder in canonical orbital ordered/disordered systems such as LaMnO$_3$ remains of great interest~\cite{Thygesen2017}, and a fuller interpretation of our findings in this 134 perovskite part doped system will form a substantial piece of future work.   

\subsection{Charge Ordering and Improper Ferroelectric Coupling}
We now give a full analysis of the average low temperature structure of HgMn$_{7}$O$_{12}$, showing that the phase transition is driven by inter-site charge transfer, followed by charge ordering on both the B and A'-sites.  Inspection of the crystal structure reveals that the formally equivalent A'-site in the rhombohedral phase is now split into three Wyckoff sites of equal multiplicity.  Furthermore, the B-sites which were split across two sites of 1:3 ratio in multiplicity, now occupy two sites of equal multiplicity.  The BVS on these 1:1 split B-sites are 3.17+ : 3.41+, and hereafter referred to as formally 3 + and 4 +.  This assignment is further validated by the evidence of Jahn-Teller elongations (2-long, 4-short bonds) present in the formal Mn$^{3+}$ state and absent in Mn$^{4+}$ (Fig.~\ref{structure}). The above description with an average B-site Mn valence of 3.5+ requires either an average formal valence state change of Mn$^{3+}$ $\rightarrow$ Mn$^{2.667+}$ on the A'-site or a valance state change of Hg$^{2+}$ $\rightarrow$  Hg$^{1+}$ to achieve charge neutrality. Hg$^{2+}$ $\rightarrow$  Hg$^{1+}$ on cooling through the phase transition can be discounted as the BVS actually shows a small increase from 1.89+ ($R\bar{3}$) to 2.12+ ($Pnn2$) on cooling, rather than a decrease expected by this scenario.  Additionally, Hg$^{+1}$ compounds are generally only stabilized by Hg dimers, and to the best of our knowledge there are no reports in the literature of stable metal oxides with undimerised  (i.e. paramagnetic) Hg(I) ions.  On the other hand, Mn$^{2+}$ in the square planar coordination geometry is known and has been reported for a number of closely related perovskite systems~\cite{Tohyama2013,Aimi2014,Zhang2013a}.  Based on crystal field stabilisation arguments, Mn$^{2+}$ in the square planar environment has a desire to restore the expected orbital degeneracy preferred by the symmetrically filled \textit{e$_g$ } and $t_{2g}$ orbitals by making its crystal field environment more regular by shortening second nearest neighbor Mn-O bond distances.  Such a shortening is evident in our low temperature refinements for Mn$^{A'}_{1}$ which has a next nearest neighbor distance of ~ 2.4 \AA ~very close to that reported in other systems with Mn$^{A' 2+}$ ~\cite{Tohyama2013,Aimi2014}, and is well below the shortest second nearest neighbor distance of 2.65 \AA ~observed for any previously reported Mn$^{A' 3+}$ system.  We hence assign a formal charge ordering of 1:2,  Mn$^{2+}$ : Mn$^{3+}$ on the A'-site.  

This scenario of charge transfer and charge ordering, is further supported by the observations of volume expansion (negative thermal expansion) on going from the high temperature rhombohedral phase to the low temperature $Pnn2$ phase, a phenomena which is observed in other 134-Perovskite charge-transfer systems ~\cite{Long2009}. Based on crystal ionic radii, the charge transfer and associated change in formal valence states is calculated to give 0.5 \AA$^{3}$ volume change per unit cell, in excellent agreement with the 0.4 \AA$^{3}$ experimentally observed (Fig. \ref{S2}). Such inter-site charge transfer is rare, but is known to act to remove unstable valence state such as in Fe$^{4+}$ containing perovskites~\cite{Long2009}.  In HgMn$_{7}$O$_{12}$, the integer inter-site transfer is between two sites which are of the same element, but have different coordinations.  To our knowledge this is not only the first known example of this, but also the first example of temperature induced charge transfer in the manganites. It seems probable that this unprecedented process acts to remove the frustration associated with the orbital disorder $R\bar{3}$ phase.

\begin{figure}
\includegraphics[width=8.9cm]{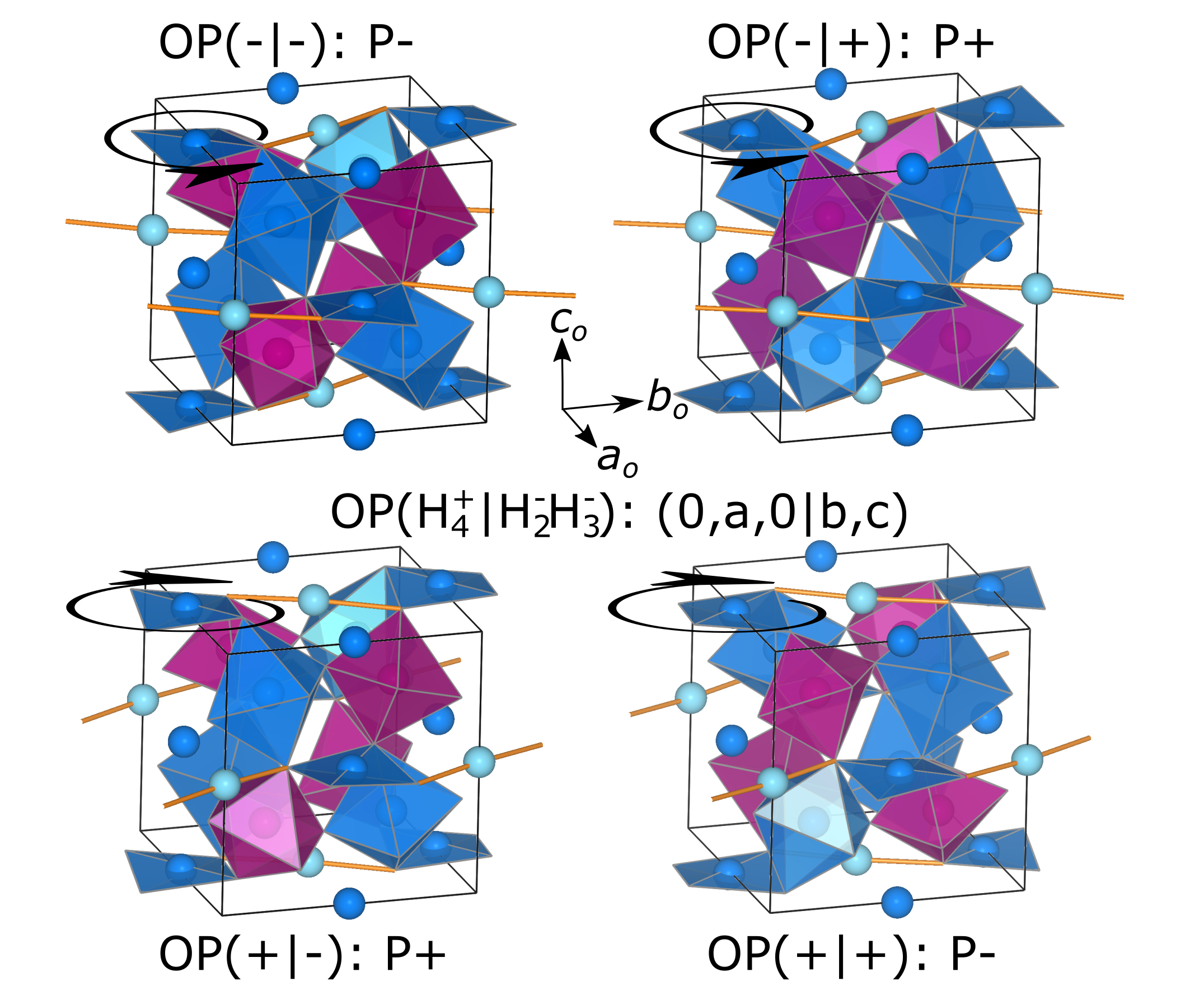}
\caption{\label{polarisation} The origin of the improper polarisation in HgMn$_{7}$O$_{12}$.  The reversal of one or other of the two primary order parameters responsible for the double-charge ordering (H$_{4}^{+}$ and H$_{2}^{-}$H$_{3}^{-}$) on the A' and B-sites is sufficient to reverse the direction of the polarisation (P), reversal of both order parameters returns the polarisation to its initial state. The short second nearest neighbor Mn-O distance ($\approx$ 2.4 \AA) that is indicative of the Mn$^{A~2+}$ state is shown in yellow (square-planar nearest neighbour coordination omitted for clarity).}
\end{figure}

The atomic displacements arising due to the charge and orbital order on the B-sites in the $Pnn2$ phase were decomposed in the $ISODISTORT$ as transforming as the zone-boundary k= [1,1,1] irreducible representations H$_{1}^{-}$ and H$_{2}^{-}$H$_{3}^{-}$ of the parent $Im\bar{3}$ phase. The irreducible representations are tabulated within $ISODISTORT$, and their characters are included in the SI (see Table S5).   A single order parameter transforming as H$_{2}^{-}$H$_{3}^{-}$ is sufficient to induce the observed orthorhombic distortion and drive the aforementioned charge and orbital order on the B-site.  However, the action of this order parameter lowers the space group symmetry only as far as $Pnnn$ and an additional primary order parameter transforming as H$_{4}^{+}$ is required to lower the symmetry to the observed $Pnn2$.  The role of the additional degrees of freedom in the $Pnn2$ model (transforming as H$_{4}^{+}$) is to help restore the orbital degeneracy on the A'-site Mn$^{2+}$ by making the coordination environment more regular and shortening the second nearest neighbor distance (Fig. \ref{polarisation}).  As both these lattice modes are necessary and sufficient to lower the symmetry of centrosymmetric $Im\bar{3}$ to the polar space group $Pnn2$, there must exist a coupling to a third secondary ferroelectric order parameter. The trilinear term responsible for the improper ferroelectric polarisation discussed in this paper is hence H$_{2}^{-}$H$_{3}^{-}$  H$_{4}^{+}$ P. We note that another possibility would be that polarisation and one of the two zone boundary modes forms the primary order parameter.  However, this scenario would lead to a  divergence in the experimentally measured dielectric constant, rather than the step-like transition (Fig. ~\ref{data}) observed in the present work which is more consistent with an improper ferroelectric mechanism ~\cite{Bousquet2008}. Unfortunately, the measurements of pyrocurrent or PE-loops in the $Pnn2$ phase at higher temperatures above the magnetic ordering transition, has been impossible due to leakage currents at these higher temperatures.  Obtaining such direct measures of the switchable nature of the crystallographically identified polar ground state is a substantial future challenge for this and other recently reported hybrid improper ferroeletrics~\cite{Benedek2011,Pitcher2015}.  Our following discussion on ferroelectric switching pathways is based on symmetry arguments alone. 

Assuming a spontaneous charge transfer and ordering on the A' and B-sites, an improper polarisation is hence always expected to arise as a secondary order parameter, and will be linearly dependent on the magnitude of the charge and orbital order on the A' and B-sites.  Additionally, the switching of the direction of the polarisation (Fig. \ref{polarisation}) will proceed by either a reversal of the checkerboard charge ordering on the B-site or by a reversal of a lattice mode that influences the A'-site charge ordering. Calculation of the polarisation from our $Pnn2$ model in the point charge approximation using average formal valence states for A and B-sites and our crystallographic coordinates, reveals a theoretical spontaneous polarisation of 8 $\mu$C cm$^{-1}$ along the crystallographic $c$-axis.  Such calculations on the topical improper ferroelectric Ca$_{3}$Ti$_{2}$O$_{7}$ (16 $\mu$C cm$^{-1}$) ~\cite{Senn2015} have found to give a reasonable estimate for the the experimental switchable polarisations (8 $\mu$C cm$^{-1}$) ~\cite{Oh2015a} and are indeed comparable to what we find here from our point charge calculations.

There has been a large amount of renewed interest in improper ferroelectric mechanisms recently, in particular in studying the coupling between two non-polar zone boundary lattice modes with octahedral tilt character, in the Ruddlesden-Popper layered perovskites, that has been discussed in the context of "hybrid improper ferroelectricity"~\cite{Benedek2011,Pitcher2015, Senn2015,Oh2015a}. Here we reveal a new mechanism by which two zone-boundary modes, implicitly dependent on orbital degrees of freedom,   combine to give polar zone-centred displacements.  These two modes should be strongly "hybridized" with each other as charge ordering on A' and B-sites most likely occurs simultaneously rather than stepwise.    

There are a few reports of other systems exhibiting polar distortions on account of charge or orbital ordering, however, many of these remain highly contentious~\cite{VandenBrink2008a, Yamauchi2014a}.  In the half-doped manganites, a combination of bond-centred and site-centred charge ordering may give rise to ferroelectricity~\cite{Efremov2004a}. More recently, it has been proposed that in cation ordered SmBaMn$_{2}$O$_{6}$, that a combination of perovskite tilting and charge ordering drives a polar ground state~\cite{Yamauchi2013}. In LuFe$_{2}$O$_{4}$, due to geometric frustration associated with charge ordering on a triangular lattice, a layered charge ordering leading with a net polarisation may occur~\cite{Ikeda2005b}. While in Fe$_{3}$O$_{4}$, the conical example of a charge ordered system, the resulting pattern of the orbital order alone breaks inversion symmetry leading to polar distortions.  However, the low temperature structure is incredibly complex and can only accurately be described by the superposition of a very large number of lattice modes~\cite{Senn2012a}.  Our present work on HgMn$_{7}$O$_{12}$ on the other hand demonstrates a compact and elegant coupling of orbital degrees of freedom directly to ferroelectric polarisation.

\subsection{Magnetoelectric Coupling}

\begin{figure}[t]
\includegraphics[width=8.9cm] {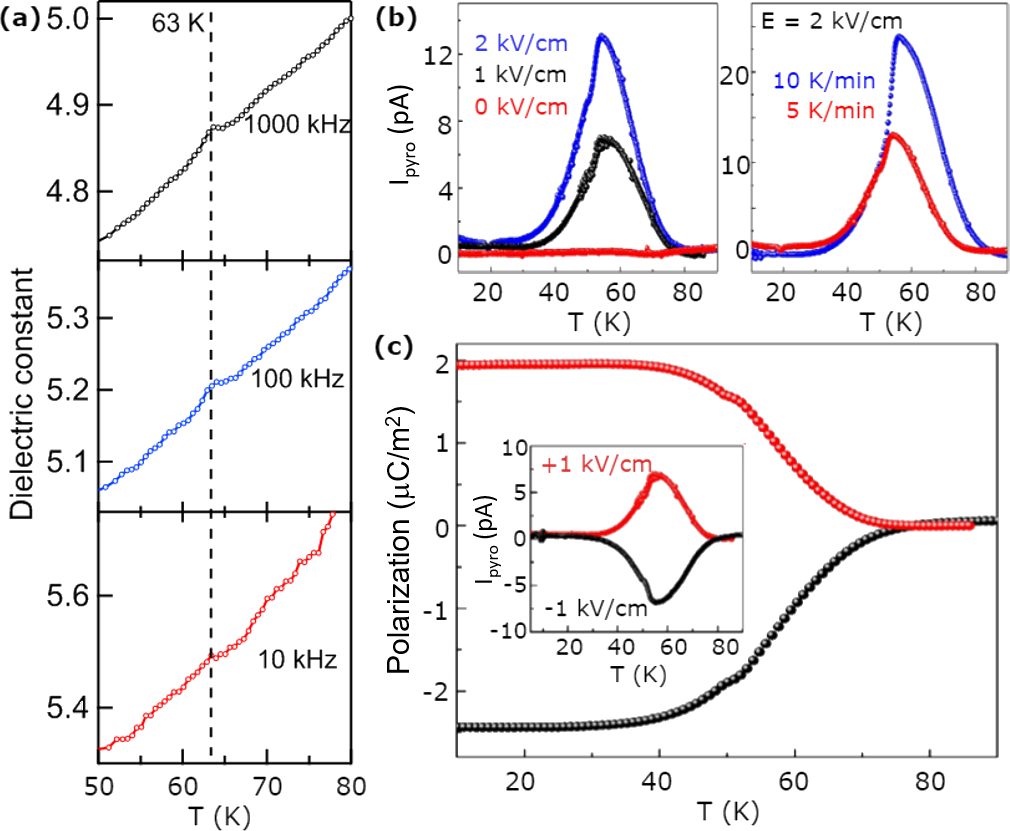}
\caption{\label{S6} Pyrocurrent results for polarization induced by the magnetic ordering transition.  (a) Anomaly at magnetic transition temperature shown in dielectric constant at various frequencies. (b) Pyrocurrent measured at various electric field and temperature ramping speeds. (c) Pyrocurrent and polarization between 10-90 K.}
\end{figure}

It is worth noting that as the underlying magnetic structure is explicitly reliant on the nature of the orbital and charge ordering in these system, any coupling between ferroelectricity and magnetism in the magnetically order ground state is expected to be very strong.  Indeed, we see experimental evidence for a  dielectric anomaly (Fig. \ref{S6}) at the magnetic ordering transition and our pyrocurrent measurements suggest an additional type-II contribution to the polarisation that is at least an order of magnitude larger  than recently measured in CaMn$_{7}$O$_{12}$ ~\cite{Terada2016}. It is important to note that our pyrocurrent results are for the polarization induced by the magnetic ordering transition only, and do not pertain to the high temperature polarization discussed in the manuscript that was calculated from the crystal structure in the point charge approximation only and is several orders of magnitude larger. A detailed understanding of the coupling terms involved in this additional low temperature contribution to the improper ferroelectric polarization will required knowledge of the low temperature magnetic structure, and this investigation will be undertaken as part of a separate communication. 

\section{Conclusion}
We have demonstrated experimentally how a relatively simple charge and orbital ordering pattern can lead to a polar structure via an improper ferroelectric mechanism that involves the coupling between two zone boundary lattice instabilities, to a third polar one. Our results represent a different realisation of schemes coupling electronic and ferroelectric order parameters together that have recently been discussed in ab initio studies on A'AB$_{2}$O$_{6}$ half-doped perovskites ~\cite{Yamauchi2013,Bristowe2015,Varignon2015}.  As the lattice instabilities that we report are directly linked to charge and orbital degrees of freedom on the A' and B-site, the coupling between electronic and ferroelectric order parameters is expected to be very strong.  This opens up a possible route for electronic control of Jahn-Teller distortions and orbitally ordered states - an appealing prospect given the numerous technologically relevant properties that depend on these fundamental orderings.

\begin{acknowledgments}
This work was supported by a joint Royal Society and Ministry of Science and Technology (MOST), Taiwan International Exchange Scheme (IE141335, 104-2911-I-002-535 and 105-2911-I-002-515).   The synchrotron beam time used in this paper was at I11 through the Diamond Light Source Block Allocation Group award “Oxford Solid State Chemistry BAG to probe composition-structure-property relationships in solids” (EE13284) and, I19 (mt13639 and mt15920), Taiwan Photon Source (2016B0091) and at ID22, European Synchrotron Source (HC2331). Neutron powder diffraction was performed at Wombat (P4864) and Echidna (MI5930). M.S.S. would like to acknowledge the Royal Commission for the Exhibition of 1851 and the Royal Society for a fellowships, and W.-T.C MOST for financial support (103-2112-M-002-022-MY3).
\end{acknowledgments}

\bibliography{HgMn7O12_bib}
\end{document}